\documentclass[10pt,pra,twocolumn,superscriptaddress,showpacs,floatfix]{revtex4}

\usepackage{graphicx}
\usepackage{amssymb}
\usepackage{times}
\usepackage{amsmath}
\usepackage{amsthm}

\usepackage{graphicx}
\usepackage{amsthm}

\begin{document}

\title{Ultra-small time-delay estimation via weak measurement technique with postselection}

\author{Chen Fang\footnote[1]{These authors contributed equally to this work.}}\affiliation
 {State Key Laboratory of Advanced Optical Communication Systems and Networks, Shanghai Key Laboratory on Navigation and Location-based Service, and Center of Quantum Information Sensing and Processing, Shanghai Jiao Tong University, Shanghai 200240, China
}

\author{Jing-Zheng Huang*\footnote[2]{jzhuang1983@sjtu.edu.cn}}\affiliation
 {State Key Laboratory of Advanced Optical Communication Systems and Networks, Shanghai Key Laboratory on Navigation and Location-based Service, and Center of Quantum Information Sensing and Processing, Shanghai Jiao Tong University, Shanghai 200240, China
}

\author{Yang Yu}\affiliation
 {State Key Laboratory of Advanced Optical Communication Systems and Networks, Shanghai Key Laboratory on Navigation and Location-based Service, and Center of Quantum Information Sensing and Processing, Shanghai Jiao Tong University, Shanghai 200240, China
}

\author{Qinzheng Li}\affiliation
 {State Key Laboratory of Advanced Optical Communication Systems and Networks, Shanghai Key Laboratory on Navigation and Location-based Service, and Center of Quantum Information Sensing and Processing, Shanghai Jiao Tong University, Shanghai 200240, China
}

\author{Guihua Zeng\footnote[3]{ghzeng@sjtu.edu.cn}}\affiliation
 {State Key Laboratory of Advanced Optical Communication Systems and Networks, Shanghai Key Laboratory on Navigation and Location-based Service, and Center of Quantum Information Sensing and Processing, Shanghai Jiao Tong University, Shanghai 200240, China
}\affiliation{College of Information Science and Technology, Northwest University, Xi¡¯an 710127, Shaanxi, China
}

\date{\today}

\begin{abstract}
Weak measurement is a novel technique for parameter estimation with higher precision. In this paper we develop a general theory for the parameter estimation based on weak measurement technique with arbitrary postselection. The previous weak value amplification model and the joint weak measurement model are two special cases in our theory. Applying the developed theory, the time-delay estimation is investigated in both theory and experiment. Experimental results shows that when the time-delay is ultra small, the joint weak measurement scheme outperforms the weak value amplification scheme, and is robust against not only the misalignment errors but also the wavelength-dependence of the optical components. These results are consistent with the theoretical predictions that has not been verified by any experiment before.
\end{abstract}

\maketitle

\section{Introduction}

As an advanced technique that provides higher sensitivity for parameter estimation, the weak  measurement scheme has attracted much attention in recent years \cite{Aharonov1988, Hosten2008,Starling2009,Viza2013,Pang2015}. Commonly, the weak measurement technique with postselection involves two physical systems, called \emph{ancillary system} and \emph{pointer}.
These systems are weakly interacted which may be described by a coupling parameter. To estimate this extremely small parameter with higher sensitivity, the ancillary system is postselected by two orthogonal states, and a followed measurement operated on the pointer system provides available information for estimating the coupling parameter.

By far, two special weak measurement techniques with postselection, i.e., the weak-value amplification (WVA) \cite{Aharonov1988} and the joint weak measurement (JWM) \cite{Strubi2013}, have been investigated. The main characteristic of these techniques is that the postselection can improve the estimation precision in the presence of technical restrictions \cite{Dressel2014,Jordan2014a,Strubi-thesis}. In the WVA scheme, the measurement result is recorded when the ancillary system is successfully postselected by one of the postselected states, which is chosen almost orthogonal to the initial state. This way leads the postselection probability to be extremely low. Using the recorded results, the coupling parameter is estimated by simply averaging the measurement results \cite{Brunner2010}. Investigation shows that the WVA is useful in suppressing some technical noises and realistic limitations, such as detector saturation, correlated noises and angular beam jitter \cite{Jordan2014a}. Subsequently, it is easier to reach the quantum standard limit with common experimental equipments in the WVA scheme \cite{Viza2014}. Moreover, in the scenarios of phase-shift and time-delay measurements, the WVA technique may provide at least 3-orders of magnitude improvement on estimation precision over the standard interferometry \cite{Brunner2010} when similar alignment errors are taken into account \cite{Li2011,Xu2013}. Of course, there are also two disadvantages, i.e., the restrictions on the estimation precision due to systematic errors and alignment errors \cite{Brunner2010,Li2011}, and the limitations on the efficiency due to the waste of large amount of unselected events after postselection operations \cite{Knee2014}.

To overcome the disadvantages in the WVA scheme, a natural way is to collect the unselected data \cite{Zhang2013,Jordan2014b}. This leads the so-called JWM scheme \cite{Strubi2013}, in which all measurement results are recorded when the events nearly equiprobably postselected by the two orthogonal postselected states. With the measurement results, the coupling parameter is estimated employing the maximum likelihood estimation method. Ref.\cite{Strubi2013} proved that by carrying out the JWM on all events, one can not only increase the efficiency but also remove systematic errors and alignment errors. Especially, the JWM technique has no ultimate precision limit \cite{Strubi2013,Strubi-thesis}, which is very useful in practice.

In this paper, we develop a general theory of weak measurement technique with arbitrary postselection. Since the involved postselections in the WVA scheme and in the JWM scheme are extremely unbalanced and almost balanced, respectively, they are actually two special cases in our general model.Our work is distinguished from the previous works that concerned in the WVA regime\cite{Lorenzo2008,Alves2015}. With the proposed general theory, the ultra-small time-delay estimation is investigated. Especially, a novel experiment for time-delay estimation making use of the weak measurement technique with postselection is firstly presented \cite{note}. Our experiment verifies exactly the theoretical predictions in Ref.\cite{Strubi-thesis}, and we find that when the time-delay is ultra small, the JWM technique outperforms the WVA technique, and is robust against not only the misalignment errors but also the wavelength-dependence of the optical components.

The paper is structured as follows. In Sec.\ref{chap:theory}, a general theory for parameter estimation based on weak measurement technique with arbitrary postselection is proposed and then is applied in the time-delay estimation. In Sec.\ref{chap:exp}, an ultra-small time-delay estimation experiment is presented. Finally, the conclusions are drawn in  Sec.\ref{chap:conclusion}.

\section{Theory}\label{chap:theory}

In this section, we develop a general theory for parameter estimation based on weak measurement technique with arbitrary postselection. Then we apply it in the ultra-small time-delay estimation.

\subsection{General Theory of Weak-Measurement-Based Parameter Estimation}

Consider a weak measurement scenario involving a two-level ancillary system prepared in state $|\varphi_i\rangle$ and a pointer with a continuous degree of freedom in state $|\phi\rangle = \int dp\phi(p)|p\rangle$, where $p$ is a continuous variable and $\phi(p)$ is the corresponding wave function.
The interaction between the system and the pointer is described by an unitary operator
\begin{equation}\label{eq:int}
\hat{U}_{int} = e^{-ig\hat{A}\hat{p}},
\end{equation}
where $\hat{A}$ acts on the system with eigenvalues of $1$ and $-1$, $\hat{p}$ acts on the pointer, and $g$ is the coupling parameter which indicates the coupling strength. This operator can be expressed as \cite{Nielsen2000}:
\begin{equation}\label{eq:interact}
\hat{U}_{int} = e^{-ig\hat{A}\hat{p}} = \cos(g\hat{p})\hat{\mathbb{I}}-i\sin(g\hat{p})\hat{A}.
\end{equation}

After the interaction, an initial product state $|\Psi\rangle = |\varphi_i\rangle\otimes|\phi\rangle$ evolves to
\begin{equation}
\begin{array}{lll}
|\Psi'\rangle &= \hat{U}_{int}|\varphi_i\rangle|\phi\rangle\\
              &= [\cos(g\hat{p})|\varphi_i\rangle-i\sin(g\hat{p})\hat{A}|\varphi_i\rangle]|\phi\rangle.
\end{array}
\end{equation}
The ancillary system is then postselected by two orthogonal states, i.e., $|\varphi_{f1}\rangle$ and $|\varphi_{f2}\rangle$, which converts the pointer state to
\begin{equation}
\begin{array}{lll}
|\phi_j\rangle &= \langle\varphi_{fj}|\Psi'\rangle\\
              &= \langle\varphi_{fj}|\varphi_i\rangle[\cos(gp)-i\sin(gp)A_{wj}]|\phi\rangle,
\end{array}
\end{equation}
where $j=1,2$, and $A_{wj} \equiv \frac{\langle\varphi_{fj}|\hat{A}|\varphi_i\rangle}{\langle\varphi_{fj}|\varphi_i\rangle}$ is the well known weak value \cite{Aharonov1988}. Consequently, the probability distribution of $p$ associated with the pointer becomes
\begin{equation}
\begin{array}{lll}\label{eq:spectrum}
P_j(p) &\equiv |\langle p|\phi_j\rangle|^2\\
          &= |\langle\varphi_{fj}|\varphi_i\rangle|^2P_0(p)\zeta_j(p,g),
\end{array}
\end{equation}
where $P_0(p) \equiv |\langle p|\phi\rangle|^2$ is the initial probability distribution of the pointer, and
$\zeta_j(p,g) \equiv \cos^2(gp) + \sin^2(gp)|A_{wj}|^2+\sin2(gp)\mathrm{Im}A_{wj}$ is the part changes the shape of the initial probability distribution.

To achieve an unbias estimate of $g$, we apply the maximum-likelihood estimation method \cite{Helstrom1976}. First, we construct the log-likelihood estimator as
\begin{equation}
\begin{array}{lll}\label{eq:L}
L(g) &= \sum_{j}\int dp Q_j(p)\log P_j(p),
\end{array}
\end{equation}
where $Q_j(p)(j=1,2)$ is the probability distribution of observing $p$ in the experiment, which in principle converges to $P_j(p)$ as the number of measured events is increased.
Then, the maximum likelihood estimate of $g$ can be achieved by solving the likelihood equation $\partial L(g)/\partial g = 0$ \cite{Shao2003}. In our case, it yields
\begin{equation}
\begin{array}{lll}\label{eq:dL}
\frac{\partial L(g)}{\partial g} &= \sum_{j}\int dp Q_j( p)\frac{\partial\log{\zeta_j(g,p)}}{\partial g}\\
                                &= \sum_{j}\int dp Q_j( p)\frac{ p[\sin2(gp)(|A_{wj}|^2-1)+2\cos2(gp)\mathrm{Im}A_{wj}]}
                                {\cos^2(gp) + \sin^2(gp)|A_{wj}|^2+\sin2(gp)\mathrm{Im}A_{wj}}\\
                                &= 0.
\end{array}
\end{equation}
As the weak measurement technique requires $gp \ll 1$ and $|A_w|gp \ll 1$ \cite{Jozsa2007}, Eq.(\ref{eq:dL}) can be simplified as
\begin{equation}
\begin{array}{lll}\label{eq:max-point}
\sum_{j}\int d p Q_j( p)\frac{-2\mathrm{Im}A_{wj} p^3g^2+(|A_{wj}|^2-1) p^2g+\mathrm{Im}A_{wj} p}
{(|A_{wj}|^2-1) p^2g^2+2\mathrm{Im}A_{wj}gp+1} = 0.
\end{array}
\end{equation}
Interestingly, Eq.(\ref{eq:max-point}) shows that prior knowledge about $P_0(p)$ is unnecessary.

Generally, the analytical solutions of Eq.(\ref{eq:max-point}) is difficult to obtain. Fortunately, under the condition of $|A_w|gp \ll 1$, Eq.(\ref{eq:max-point}) can be approximately simplified to a quartic equation
\begin{equation}
Ag^4 + Bg^3 + Cg^2 + Dg +E = 0,
\end{equation}
where\\
$A=\sum_{j}\int d p Q_j(p) p^5[2(|A_{wj}|^2-1)\mathrm{Im}A_{wj}]$,\\
$B=\sum_{j}\int d p Q_j(p) p^4[4(\mathrm{Im}A_{wj})^2-(|A_{wj}|^2-1)^2]$,\\
$C=\sum_{j}\int d p Q_j(p) p^3[\mathrm{Im}A_{wj}(1-3|A_{wj}|^2]$,\\
$D=\sum_{j}\int d p Q_j(p) p^2[(|A_{wj}|^2-1)-2\mathrm{Im}A_{wj}]$, and\\
$E=\sum_{j}\int d p Q_j(p) p\mathrm{Im}A_{wj}$.\\
As $g$ is extremely small, one may obtain the first-order approximate solution
\begin{equation}\label{eq:first-order}
g^{(1)}_{est} = -\frac{E}{D}
              = \frac{\sum \mathrm{Im}A_{wj}\langle p\rangle_j}{\sum[2(\mathrm{Im}A_{wj})^2-|A_{wj}|^2+1]\langle p^2\rangle_j},
\end{equation}
where $\langle p\rangle_j = \int dpQ_j(p)p$ and $\langle p^2\rangle_j = \int dpQ_j(p)p^2$. Eq.(\ref{eq:first-order}) provides an appropriate estimate of $g$ with very high precision.

\subsection{Time-delay estimation}\label{subsec:time-delay}
\begin{figure}[!h]\center
\resizebox{9cm}{!}{
\includegraphics{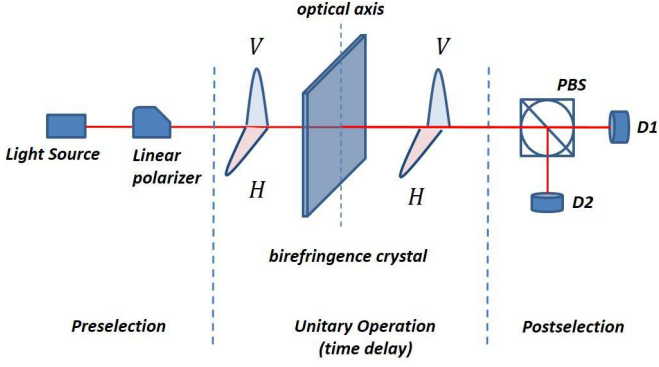}}
\caption{(Color online). Time-delay estimation. Photons emitted from a light source is modulated to be a linear superposition state of two orthogonal polarizations, $H$ and $V$, by a linear polarizer. After passing through a thin birefringent crystal, a small time delay is involved between the two polarizations. The followed circular polarization beam splitter(PBS) postselects two polarization states, and two spectrometers (D1 and D2) are placed to measure the spectrum of the output lights.}\label{fig:time-delay}
\end{figure}

As an application of our developed theory, we consider the time-delay estimation described in Fig.\ref{fig:time-delay}. In this scheme, an ultra small time delay of $\tau$ between two orthogonal polarizations, which are denoted as horizontal ($H$) and vertical ($V$), is involved by a thin birefringent crystal. The corresponding unitary operator is given by
\begin{equation}\label{eq:time-delay}
\hat{U}_{int}(\tau) = e^{-i\tau\hat{\sigma}_Z\omega},
\end{equation}
where $\hat{\sigma}_Z = |H\rangle\langle H|-|V\rangle\langle V|$ is the $Z$ operator acting on the polarization states and $\omega$ represents the frequencies of the photons. In this scenario the parameters $g$ and $p$ in the general model are replaced by time delay $\tau$ and frequency $\omega$, respectively, and the photon polarizations and frequencies are chosen as the ancillary system variable and the pointer variable, respectively.

Given an initial state of $|\varphi_i\rangle = \frac{|H\rangle + |V\rangle}{\sqrt{2}}$ and post-selected states of $|\varphi_{f1}\rangle = \frac{|H\rangle + e^{i\phi}|V\rangle}{\sqrt{2}}$ and $|\varphi_{f2}\rangle = \frac{|H\rangle - e^{i\phi}|V\rangle}{\sqrt{2}}$, we get the weak values
\begin{equation}
\begin{array}{lll}\label{eq:weak-value}
A_{w1} = \frac{\langle \varphi_{f1}|\hat{A}|\varphi_i\rangle}{\langle \varphi_{f1}|\varphi_i\rangle} = \frac{1-e^{-i\phi}}{1+e^{-i\phi}} = i\tan(\frac{\phi}{2}) \\
A_{w2} = \frac{\langle \varphi_{f2}|\hat{A}|\varphi_i\rangle}{\langle \varphi_{f2}|\varphi_i\rangle} = \frac{1+e^{-i\phi}}{1-e^{-i\phi}} = -i\cot(\frac{\phi}{2}).
\end{array}
\end{equation}
The postselection probabilities are given by
\begin{equation}
\begin{array}{lll}
P_{f1} &= |\langle\varphi_{f1}|\varphi_i\rangle|^2\int d\omega P_0(\omega)\zeta_1(\omega,\tau)\\
       &= \cos^2\frac{\phi}{2}[1+(\tan^2\frac{\phi}{2}-1)\omega_0^2\tau^2+2\tan\frac{\phi}{2}\omega_0\tau];\\
P_{f2} &= |\langle\varphi_{f2}|\varphi_i\rangle|^2\int d\omega P_0(\omega)\zeta_2(\omega,\tau)\\
       &= \sin^2\frac{\phi}{2}[1+(\cot^2\frac{\phi}{2}-1)\omega_0^2\tau^2-2\cot\frac{\phi}{2}\omega_0\tau],
\end{array}
\end{equation}
where $\omega_0 \equiv \int P_0(\omega)\omega d\omega$ represents the initial average frequency of light before the interaction.
In practice, $P_{f1}$ and $P_{f2}$ can be estimated from experimental data by $\int Q_1(\omega)d\omega$ and $\int Q_2(\omega)d\omega$, respectively.
%

Inserting Eq.(\ref{eq:weak-value}) into Eq.(\ref{eq:first-order}) gives
\begin{equation}
\begin{array}{lll}\label{eq:time-delay-approx}
\tau^{(1)}_{est} =\frac{\sin^2\frac{\phi}{2}\langle\omega\rangle_1-\cos^2\frac{\phi}{2}\langle\omega\rangle_2}{\tan\frac{\phi}{2}\langle\omega^2\rangle_1+\cot\frac{\phi}{2}\langle\omega^2\rangle_2}.
\end{array}
\end{equation}
We note that estimating $\tau$ with Eq.(\ref{eq:time-delay-approx}) requires prior knowledge on $\phi$, which is, however, not necessary in high precision for achieving high signal-to-noise ratio\cite{Strubi2013}. Further discussions can be found in Sec.II.C.
\begin{figure}[!h]\center
\resizebox{6cm}{!}{
\includegraphics{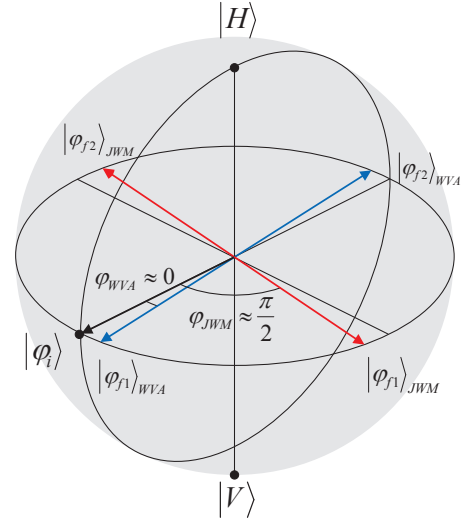}}
\caption{(Color online). Representation on the Bloch sphere of the pre-selected and post-selected states, where the subscripts WVA and JWM stand for two schemes.}
\label{fig:bloch}
\end{figure}

Now we consider applications of the proposed theory model in two specific schemes, i.e., the JWM scheme and the WVA scheme. In the JWM scheme, the light intensities detected by two spectrometers are nearly equal, which requires $\phi \approx \frac{\pi}{2}$, the pre-selected and post-selected states can be seen in Fig.\ref{fig:bloch}. The postselection probabilities of two output ports are
\begin{equation}
\begin{array}{lll}\label{eq:p-jw}
P_{f1} &\approx \cos^2\frac{\phi}{2}(1+2\tan\frac{\phi}{2}\omega_0\tau).\\
P_{f2} &\approx \sin^2\frac{\phi}{2}(1-2\cot\frac{\phi}{2}\omega_0\tau).
\end{array}
\end{equation}
In the JWM scheme, the weak value is not anomalous, thus one can not observe an significant shift on the pointer's wave function, as it is the case in WVA. In fact, the measured shift is existed in the probability difference\cite{MR2016}.

Then, Eq.(\ref{eq:time-delay-approx}) can be simplified as
\begin{equation}
\begin{array}{lll}\label{eq:jw}
\tau_{jw} \approx \frac{P_{f1}\langle\omega\rangle_2-P_{f2}\langle\omega\rangle_1}{\Delta\omega^2}.
\end{array}
\end{equation}
Employing Eq.(\ref{eq:jw}) the time-delay $\tau$ can be estimated without being precisely aware of $\phi$.
We note that this result is conflict with the formula given in \cite{Strubi2013}, which is
$$\tau = \frac{1}{4}( \frac{\langle\omega\rangle_2-\langle\omega\rangle_1}{\Delta\omega^2} - \frac{P_{f2}-P_{f1}}{\Delta\omega}).$$
In Sec.\ref{chap:exp} we will further verify the correctness of our formula by experiment.

Next we consider the application of the proposed theory model in WVA technique. In the WVA scheme, most of the light output from port 1, which requires $\phi \approx 0$ as shown in Fig.\ref{fig:bloch}. The postselection probabilities of the two output ports are
\begin{equation}
\begin{array}{lll}\label{eq:p-wva}
P_{f1} &\approx  (1-\frac{\phi^2}{4})[1+\phi\omega_0\tau].\\
P_{f2} &\approx \frac{\phi^2}{4}[1-\frac{4}{\phi^2}\omega_0\tau].
\end{array}
\end{equation}
In this case, Eq.(\ref{eq:time-delay-approx}) gives
\begin{equation}
\begin{array}{lll}\label{eq:wva}
\tau_{wva}= \frac{\frac{\phi}{2}(\overline{\langle\omega\rangle}_1-\overline{\langle\omega\rangle}_2)}{\overline{\langle\omega^2\rangle}_1+\overline{\langle\omega^2\rangle}_2-2\omega_0\overline{\langle\omega\rangle}_2},
\end{array}
\end{equation}
where
$$\overline{\langle\omega\rangle}_j \equiv \frac{ \int Q_j(\omega)\omega d\omega}{\int Q_j(\omega)d\omega} = \langle\omega\rangle_j/P_{fj},$$
and
$$\overline{\langle\omega^2\rangle}_j \equiv \frac{ \int Q_j(\omega)\omega^2 d\omega}{\int Q_j(\omega)d\omega} = \langle\omega^2\rangle_j/P_{fj}.$$

In Eq.(\ref{eq:wva}), we have $\overline{\langle\omega\rangle}_1-\overline{\langle\omega\rangle}_2 \approx \omega_0 - \overline{\langle\omega\rangle}_2 \equiv \delta\omega$ and  $\overline{\langle\omega^2\rangle}_1+\overline{\langle\omega^2\rangle}_2-2\omega_0\overline{\langle\omega\rangle}_2 \approx 2\langle\Delta\omega^2\rangle$. Reminding that $A_{w2} = -i\cot(\frac{\phi}{2}) \approx -i\frac{2}{\phi}$, we get
\begin{equation}
\begin{array}{lll}
\delta\omega \approx 2\tau_{est}ImA_{w2}\langle\Delta\omega^2\rangle,
\end{array}
\end{equation}
which is the familiar formula derived in the weak value amplification scheme \cite{Li2011}.

\section{Experiment}\label{chap:exp}

In this section, we perform an experiment to demonstrate the time delay estimation based on the theory model presented in Sec.\ref{subsec:time-delay}. Making use of the experimental results, we study the performances of the JWM and WVA schemes, so that we can verify the theoretical predictions about the advantages of JWM presented in Ref.\cite{Strubi2013}.

\subsection{Experimental Setup}

\begin{figure}[!h]\center
\resizebox{9cm}{!}{
\includegraphics{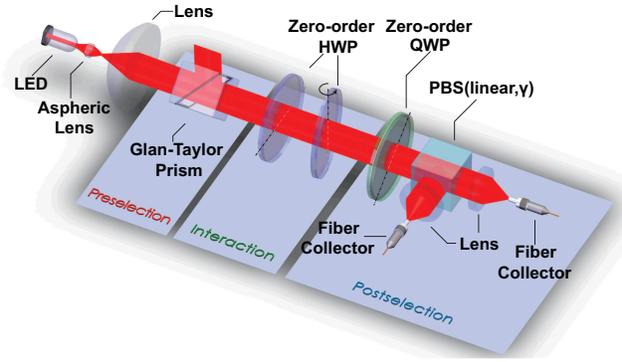}}
\caption{(Color online). Experimental setup of the time-delay estimation. Photons emitted from a light emitted diode (LED) with center wavelength of 780nm and line width of $\Delta\lambda = 17.6 nm$ are collimated by an aspheric lens and a convex plane lens then enter the Glan-Taylor prism, which preselects the polarization state $|\varphi_i\rangle = \frac{1}{\sqrt{2}}(|H\rangle + |V\rangle)$. After that, the light passes through two compound binary zero-order half-wave plates(HWPs) with their optical axes(OAs) perpendicular to each other and at $45^{\circ}$ to the polarization of the prism.The second HWP is pivoted by a small angle $\theta$ around its OA. Postselction is realized by a combination of a compound binary zero-order quarter-wave plate(QWP) and a polarization beam splitter(PBS).The photons are split into two orthogonal polarized beams and focused into optical fibers by lens, then sent to two spectrometers with sampling resolution of 0.1 nm and range of 690-900 nm.
}\label{fig:experiment}
\end{figure}

The experimental setup is described in Fig.\ref{fig:experiment}, in which a thin birefringent crystal is used to induce the time delay. We note that it is very difficult to produce and manipulate this kind of plate in practice. Follow the idea in Ref.\cite{Xu2013}, we use a double-plate system with placing their optical axes(OAs) perpendicularly to each other to equivalently realize this effect. Restricted by the laboratory condition, we use the binary compound zero-order half-wave plates (HWPs) in our experiment. These plates are built by combing two multi-order wave plates and aligning the fast axis of one plate with the slow axis of the other to obtain the zero-order phase delay. In contrast to the true zero-order plates used in \cite{Xu2013}, they are insensitive to the temperature changes. We place both plates perpendicularly to the light velocity to cancel their phase delay. The delay is realized by pivoting the second plate around its OA by a tiny angle $\theta$. This pivot increases the optical path of the second HWP, which makes the double-HWPs system equivalent to a very thin plate of the same material with the OA orienting as the tilted one. The relationship between the time delay $\tau$ and the tilt angle $\theta$ is
\begin{equation}
\begin{array}{lll}\label{eq:tau}
\tau = \frac{(n_e-n_o)h\theta^2}{2c\lambda n^2},
\end{array}
\end{equation}
where $n_e$, $n_o$ and $n$ are the refractive indexes of quartz for extraordinary light, ordinary light and average, respectively, h is the thickness of the plate, c is the light speed in vacuum and $\lambda$ is the wavelength of the light.
For detailed derivations, see Appendix. A and Ref.\cite{Li2002}.

Finally, in order to finish the postselection, the photons sequentially enter a quarter-wave plate (QWP) and a polarization beam splitter (PBS).
Originally, the OA of the QWP is perpendicular to the axis of the Glan-Tylor prism and the PBS splits the photons to polarization states $|H\rangle$ and $|V\rangle$. And then, the PBS is rotated by an angle of $\gamma = \pi/4 - \frac{\phi}{2}$ according to the GT prism's OA orientation to postselect final states of $|\varphi_{f1}\rangle = \cos\gamma|H\rangle + \sin\gamma|V\rangle$ and $|\varphi_{f2}\rangle = \sin\gamma|H\rangle - \cos\gamma|V\rangle$.

In our setting, when $\phi \approx \frac{\pi}{2}$, the light intensities go out from the two output ports are nearly the same, which corresponds to the JWM proposal\cite{Strubi2013}. In contrast when $\phi \approx 0$, almost all of the light go out from one output port, which corresponds to the weak-value amplification scheme. In the following, we will present the experimental results of these two schemes to verify the theory derived in Sec.\ref{chap:theory}.

\subsection{Joint Weak Measurement v.s. Weak Value Amplification}\label{subsec:balanced}

\begin{figure}[!h]\center
\resizebox{9cm}{!}{
\includegraphics{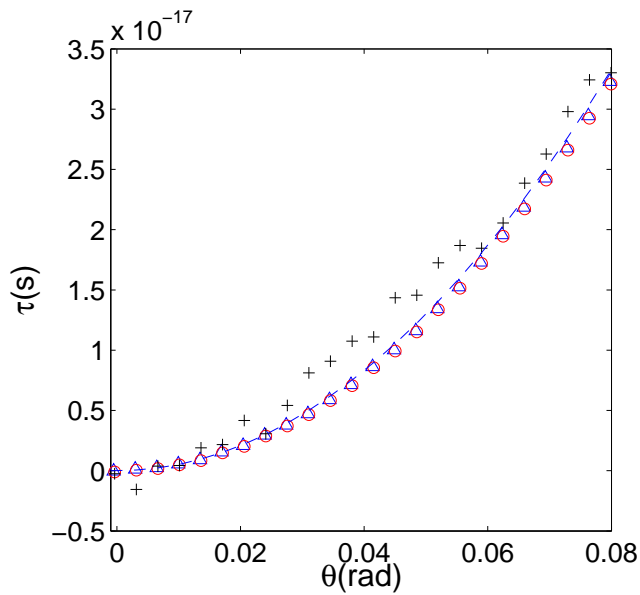}}
\caption{(Color online.) Relation of $\theta$ and $\tau$ in the joint weak measurement scheme. Blue line presents the theoretical prediction from Eq.(\ref{eq:tau}), blue triangles, red circles and black crosses are estimates of $\tau$ from experimental data, calculated by the most rigorous equation [Eq.(\ref{eq:max-point})], the first-order approximation [Eq.(\ref{eq:time-delay-approx})] and the simplified approximation [Eq.(\ref{eq:jw})] respectively.
}\label{fig:jw-complete}
\end{figure}

%
%
%
Firstly, we consider the JWM technique where the postselection is almost balanced. In this scheme, we set $\phi \approx (\frac{\pi}{2}+0.071)~rad$ (by rotating the PBS with an angle of $-0.071~ rad$), and the postselection probabilities for each output port can be calculated by Eq.(\ref{eq:p-jw}).
In Fig.\ref{fig:jw-complete}, according to the tilted angle we made, different estimations about the time-delay are shown, where the theoretical predictions drawn by blue line are calculated by Eq.(\ref{eq:tau}). Among these estimations, blue triangles are the complete solutions to Eq.(\ref{eq:max-point}), when red circles and black crosses are the approximate solutions given by Eq.(\ref{eq:time-delay-approx}) and Eq.(\ref{eq:jw}) respectively.
Compared to Eq.(\ref{eq:time-delay-approx}), Eq.(\ref{eq:jw}) does not require the prior knowledge about $\phi$, this parameter is estimated by the light intensity difference of the two output ports. However Eq.(\ref{eq:jw}) provides an estimation with less accuracy, because the environment effects, such as the fluctuation of the temperature, make the output intensity of the LED unstable  .
%
%

%
%
%
%
%

\begin{figure}[!h]\center
\resizebox{9cm}{!}{
\includegraphics{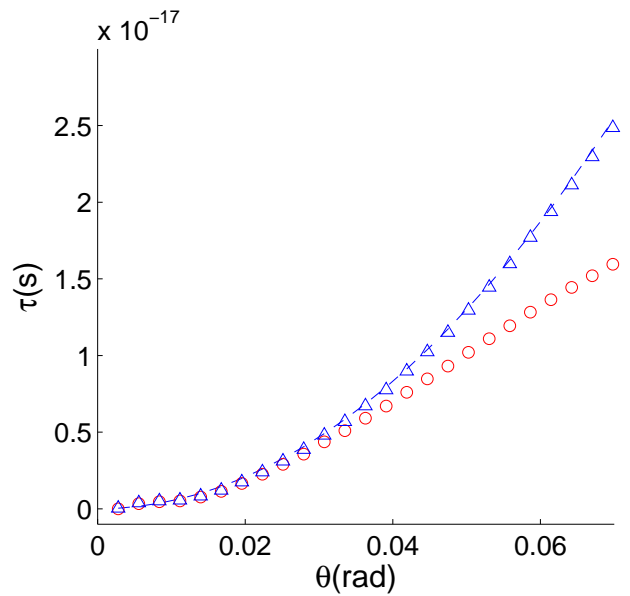}}
\caption{(Color online.) Relation of $\theta$ and $\tau$ in the weak value amplification scheme. Blue line presents the theoretical prediction from Eq.(\ref{eq:tau}), blue triangles and red circles are the estimates of $\tau$ from experimental data, calculated by the most rigorous equation [Eq.(\ref{eq:max-point})] and the first-order approximation [Eq.(\ref{eq:time-delay-approx})] respectively.
}\label{fig:wva-complete}
\end{figure}

Then, we consider the WVA technique where the postselection is extremely unbalanced. In this scheme, we set $\phi \approx 0.03$ (by rotating the PBS for $\pi/4-0.03$ $rad$), and the postselection probabilities for each output port, which can be calculated by Eq.(\ref{eq:p-wva}), are extremely unbalanced. In Fig.\ref{fig:wva-complete}, according to the tilted angle we made different estimations about the time-delay are shown, where the theoretical predictions drawn by blue line are calculated by Eq.(\ref{eq:tau}). Among these estimations, blue triangles are the complete solutions to Eq.(\ref{eq:max-point}), when red circles are the approximate solutions given by Eq.(\ref{eq:time-delay-approx}).
%

We can find that the red circles fit well with the theory when $\tau$ is less than $\ensuremath{0.5\times10^{-17}}$ but the deviation increases along with the growing $\tau$, because the effect of wavelength dependency of the quarter-wave plate was not taken into account in Eq.(\ref{eq:wva}) (see Appendix. B for details). For more discussions about this phenomenon, we refer the readers to Refs.\cite{Koike2011,Zhou2014,Chen2015}.
%
%
%

\subsection{Discussions}


%
Utilizing the imaginary part of weak value provides a large amplification factor for parameter estimation. For this purpose, a quarter-wave plate for adding a circular component on the polarization is required. However, as is pointed out in \cite{Xu2013}, the quarter-wave plate involves large uncertainty to the polarization that is difficult to compensate, because of the wide spectrum of the light source.

There are two possible ways to solve this problem: get rid of the quarter-wave plate with the price of much smaller weak value, or apply joint weak measurement method to suppress the wavelength-dependence effect.

In our experimental setting (see Fig.\ref{fig:experiment}), removing the quarter-wave plate reverts a similar setup as that in \cite{Xu2013}. Alternatively, we can collect the data from both output sides and apply Eq.(\ref{eq:first-order}) for parameter estimation. When the condition $|A_w|\omega_0\tau \ll 1$ is satisfied, the estimation fits quite well to the theoretical prediction (see Fig.\ref{fig:wva-complete}).


In \cite{Xu2013}, the uncertainty of measuring $\alpha$ (equals $\omega_0\tau$ in our paper) depends on the spectrometer resolution $\Delta(\delta\lambda)$ and the postselection parameter $\beta$ (equals $2\gamma$ in our paper). The uncertainty of $\alpha$, which is denoted as $\Delta\alpha$, can be derived as\cite{Xu2013}:
\begin{equation}
\begin{array}{lll}\label{eq:uncertainty}
\Delta\alpha = \frac{\lambda_0}{4\Delta\lambda^2}\cdot\frac{(\alpha^2+\beta^2)^2}{\alpha\beta^2}\cdot\Delta(\delta\lambda).
\end{array}
\end{equation}
According to this formula, the optimal precision can be achieved when $\beta \simeq \alpha$, but it is very impractical because the value of $\alpha$ is unknown before it is measured. Alternatively, the authors pointed out that for measuring small $\alpha$ one can set $\beta = 0$ and obtain $\Delta\alpha \simeq 0.1\alpha$ when $\Delta(\delta\lambda) = 0.1 nm$. However, an important factor, the misalignment of $\beta$, had not been considered. In fact, it sets a limitation on the minimum detectable phase when using weak value amplification\cite{Brunner2010}. Assume that the actual value of $\beta$ is $\epsilon$ when we set it to be 0, and consider $\alpha_{min} = \Delta\alpha$ as the minimum detectable value of the measured phase, we can derived from Eq.(\ref{eq:uncertainty}) that
\begin{equation}
\begin{array}{lll}\label{eq:alphamin}
\alpha_{min} = \sqrt{\frac{\sqrt{1-4C}-2C+1}{2C}}\epsilon,
\end{array}
\end{equation}
where we define $C \equiv \frac{\lambda_0}{4\Delta\lambda^2}\Delta(\delta\lambda)$ for simplicity.
By setting $\Delta(\delta\lambda)=0.1 nm$, which is assumed in \cite{Xu2013} and practically applied in our experiment, we get $\alpha_{min} \simeq 3.7 \epsilon$.

\begin{figure}[!h]\center
\resizebox{9cm}{!}{
\includegraphics{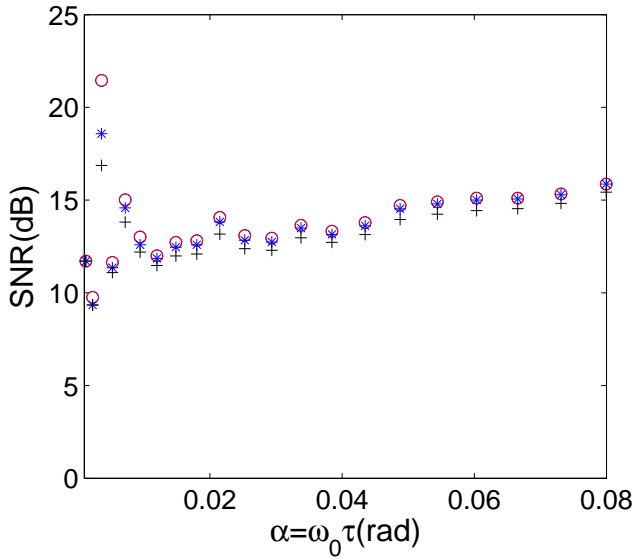}}
\caption{Signal-to-noise ratio in the time-delay estimation experiment with joint weak measurement. Red circles: $\phi=0.03$; Blue stars: $\phi=0.05$; Black crosses: $\phi=0.08$.}\label{fig:snr-uncertain}
\end{figure}

On the other hand, by using the joint weak measurement scheme proposed in \cite{Strubi2013} and demonstrated in Sec.\ref{subsec:balanced}, the uncertainty of measured phase is insensitive to the alignment error.
In our experiment, we set $\phi=0.03$. The exact value of $\phi$ can be estimated by maximizing Eq.(\ref{eq:max-point}), but it is not necessary to do so for achieving high precision.
In Fig.\ref{fig:snr-uncertain} we calculate the signal-to-noise ratio(SNR) when estimating $\tau$ by Eq.(\ref{eq:first-order}) and setting 
$\phi=0.03$, $0.05$ and $0.08$ rad respectively. The SNR drops slightly even when our prediction about $\phi$ has a deviation of $0.05$ rad, and remains higher than 10 dB for $\alpha > 0.002$. It shows that the estimation results are insensitive to $\phi$ in balanced postselection scheme.
In contrast, for weak value amplification in unbalanced postselection scheme, for a typical deviation of $\epsilon = 0.0027$\cite{Xu2013}, the SNR drops down to 0 dB when $\alpha = 0.01$, much worse than applying the joint weak measurement method.
%

As is predicted by theory\cite{Strubi2013} and presented in Fig.\ref{fig:snr-uncertain}, the SNR of joint weak measurement scheme remains stable when $\alpha$ grows, because the estimation error increases along with the signal. As  in the weak value amplification scheme the SNR increases along with the signal\cite{Starling2010}, it implies that at some point the weak value amplification scheme will achieve higher SNR than the joint weak measurement scheme.
In that case, one can adopt the weak value amplification scheme for better performance, however the harm of the QWP must be overcome by, e.g., applying a broadband wave plate. This is beyond the scope of our current work.

Finally, it is useful to compared the JWM and WVA with other schemes for practical purpose. In Ref.\cite{Brunner2010} the authors compared the WVA and standard interferometric scheme for phase shift estimation, and proved that when similar alignment errors are taken into account, the estimation precision of WVA is at least 3-orders of magnitude higher than the standard interferometric scheme. Combining with the above discussions, it is reasonable to expect that JWM has the highest precision for estimating extremely small time delay among these three schemes.

\section{Conclusion}\label{chap:conclusion}

In summary, we present a general theory for parameter estimation based on weak measurement with arbitrary postselection. Applying this theory, we study the time-delay estimation in both theory and experiment, with especial interest in two specific schemes, i.e., the WVA scheme and the JWM scheme. Through the experimental results, we find that the JWM scheme outperforms the WVA scheme when the time-delay is ultra small. These results support the theoretical predictions presented in Ref.\cite{Strubi2013} that has not been verified by any experiment before. Furthermore, the JWM technique is robust against not only the misalignment errors but also the wavelength-dependence of the optical components, which appears to be an advanced parameter estimation approach for achieving higher precision with convenient laboratory equipments.

\begin{acknowledgments}
This work was supported by the National Natural Science Foundation of China (Grants No. 61170228£¬No. 61332019£¬ 61471239), and the Hi-Tech Research and Development Program of China (Grant No: 2013AA122901).
\end{acknowledgments}


\appendix

\section{Phase retardation of compound binary zero-order wave-plate}

When light propagates through a uniaxial crystal plate, the phase delay between the ordinary light and extraordinary light is,
 \begin{equation}
\begin{array}{lll}\label{eq:delta}\\
\delta=\frac{2\pi}{\lambda}(n_e-n_o)L\sin^2\eta,\\
\end{array}
\end{equation}
where $\lambda$ is the wavelength,$n_o$ and $n_e$ stand for refractive index of ordinary and extraordinary light,respectively.$L=\frac{h}{\cos\eta_i}$ means the path length propagating in the birefringent crystal. $\eta$ is the angle between the normal line and the OA,$\eta_i$ is the angle between the light velocity in the crystal and the normal line.And $h$ is the thickness of the plate.
\begin{figure}[!h]\center
\resizebox{9cm}{!}{
\includegraphics{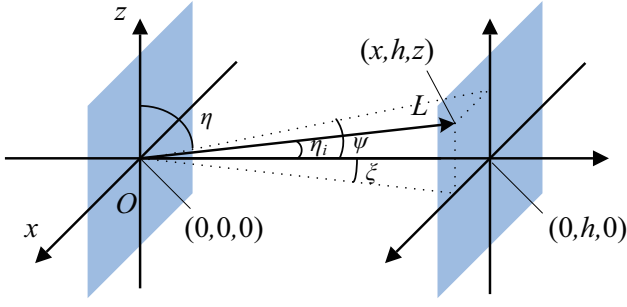}}
\caption{(Color online). Coordinate system of single birefringent waveplate.}\label{fig:coordinate1}
\end{figure}
For simplicity,we establish a Cartesian coordinates showed in Fig.\ref{fig:coordinate1} to calculate $C=L\sin^2\eta$ in the crystal which we call the effective path length in the next. The origin is coincident with the point of incidence.X-Z plane is on the surface of the plate.And the OA is paralleled with Z-axis.
It's obvious that $L^2=x^2+y^2+z^2,L^2\sin^2\eta=x^2+y^2$,and
 \begin{equation}
\begin{array}{lll}\label{eq:eff_length}\\
C=L\sin^2\eta=\frac{x^2+y^2}{\sqrt{x^2+y^2+z^2}}.\\
\end{array}
\end{equation}
The exit point is$(x,h,z)$.And the effective path length could be obtain by replacing $h$ by $y$.In our experiment, the pivot is around the OA, we define the azimuth angle and elevation angle as $\xi$ and $\psi$ respectively, so
 \begin{subequations}\label{eq:length}\\
 \begin{equation}
 \begin{array}{lll}\\
x^2+h^2=h^2/\cos^2\xi,\\
z^2+h^2=h^2/\cos^2\psi,\\
\end{array}
\end{equation}
and we could deduce that
 \begin{equation}
 \begin{array}{lll}\
x^2+y^2+z^2=\frac{h^2(1-\sin^2\xi\sin^2\psi)}{\cos^2\xi\cos^22\psi)},\\
\end{array}
\end{equation}
\end{subequations}
inserting Eq.\ref{eq:eff_length} and Eq.\ref{eq:length} into Eq.\ref{eq:delta}, we could get the general expression of the phase retardance of oblique incidence on a crystal plate
 \begin{equation}
 \begin{array}{lll}\
\delta=\frac{2\pi}{\lambda}(n_e-n_o)C=\frac{2\pi(n_e-n_o)h\cos\psi}{\sqrt{\cos\xi(1-\sin^2\xi\sin^2\psi)}}.\\
\end{array}
\end{equation}

The compound binary plate is built by combining two multi-order wave plates to obtain an zero-order phase delay by aligning the fast axis of one plate with the slow axis of the other.The retardance could be expressed as
\begin{equation}
 \begin{array}{lll}\
\delta=\frac{2\pi}{\lambda}(n_e-n_o)(h_1-h_2).\\
\end{array}
\end{equation}

\begin{figure}[!h]\center
\resizebox{9cm}{!}{
\includegraphics{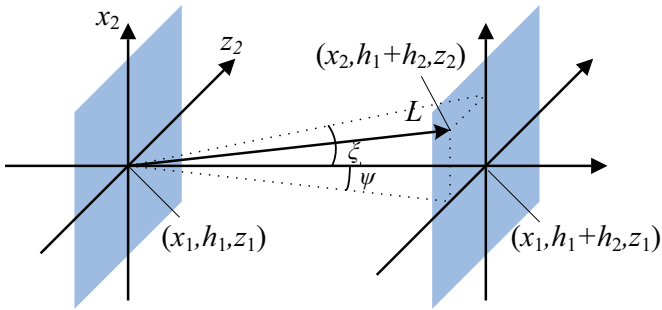}}
\caption{(Color online). Coordinate system of the second plate in a compound wave plate.}\label{fig:coordinate2}
\end{figure}

Fig.\ref{fig:coordinate2} represent the Cartesian coordinates when light go through the second plate while the Fig.\ref{fig:coordinate1} still shows the incidence on the first plate.The origin overlaps with the exit point of the first plate.Since the OA of the second plate is perpendicular to the first one's, the coordinate system is rotating by $90^\circ$.  Let the subscripts denote the two plates.
We neglect the refringence between the two plates,and suppose that $\xi$ in Fig.\ref{fig:coordinate2} is approximately equal to $\psi$ in Fig.\ref{fig:coordinate1}, we can derive that
 \begin{equation}
 \begin{array}{lll}\
z_1/h_1=x_2/h_2,\\
x_1/h_1=-z_2/h_2.\\
\end{array}
\end{equation}
The phase delay of the compound plate is
 \begin{equation}
 \begin{array}{lll}\
\delta=\frac{2\pi}{\lambda}(n_e-n_o)(C1-C2),\\
\end{array}
\end{equation}
where
 \begin{equation}
 \begin{array}{lll}\
C1=\frac{x_1^2+h_1^2}{\sqrt{x_1^2+h_1^2+z_1^2}},\\
C2=\frac{h2(x_1^2+h_1^2)}{h1\sqrt{x_1^2+h_1^2+z_1^2}},\\
\end{array}
\end{equation}

combining with Eq.\ref{eq:eff_length} and taking account of the relationship of those angles, we found the general expression of the phase retardance of the compound plate:
  \begin{equation}
 \begin{array}{lll}\label{eq:CPdelta}\\
\delta=\frac{2\pi(n_e-n_o)(h_1\cos\psi/\cos\xi-h_2\cos\xi/\cos\psi)}{\lambda\sqrt{1-sin^2\xi\sin^2\psi}}.\\
\end{array}
\end{equation}
when $\psi=0$ and $\xi$ is small, Eq.\ref{eq:CPdelta} becomes
\begin{subequations}\label{eq:approxdelta}\\
\begin{equation}
\begin{array}{lll}
\delta &= \frac{2\pi(n_e-n_o)}{\lambda}(h_1/\cos\xi-h_2\cos\xi)\\
        &\approx \frac{2\pi(n_e-n_o)}{\lambda}[(h_1-h_2)+\xi^2(h_1+h_2)/2],\\
\end{array}
\end{equation}
and when $\xi=0$ and $\psi$ is small, Eq.\ref{eq:CPdelta} approximates to
\begin{equation}
\begin{array}{lll}
\delta &= \frac{2\pi(n_e-n_o)}{\lambda}(h_1\cos\psi-h_2/cos\psi)\\
        &\approx \frac{2\pi(n_e-n_o)}{\lambda}[(h_1-h_2)-\psi^2(h_1+h_2)/2].\\
\end{array}
\end{equation}
\end{subequations}
In Eq.\ref{eq:approxdelta}, the 1st term is the plate's retardance of vertical incidence,the 2nd term means the phase delay of oblique incidence.Considering the relationship between the incidence angle $\theta,\psi$ and $\xi$, we could rewrite the 2nd term as
\begin{equation}
\begin{array}{lll}
\Delta\delta=\frac{\pm\pi(n_e-n_o)(h_1+h_2)\theta^2}{\lambda n^2}\\
\end{array}
\end{equation}
when $\psi=0$ it has the positive value and the negative value could be obtain while $\xi=0$.
In the range of our LED spectrum ,$n_e-n_o$ is almost constant, in other words, the optical length in the crystal is unsensitive to the variation of wavelength.With the cancelling effect of the two waveplates in our experiment, the time delay induced by the pivot is
\begin{equation}
\begin{array}{lll}
\tau=\frac{\Delta\delta}{2\pi}\frac{\lambda}{c}=\frac{\pm(n_e-n_o)(h_1+h_2)\theta^2}{2c\lambda n^2}\\
\end{array}
\end{equation}
where $c$ is the the speed of light in vacuum.

\section{Effect of wavelength-dependent quarter-wave plate}\label{}

The postselection in our experiment consists of a quarter-wave plate and a linear polarization beam splitter(PBS). For using ideal quarter-wave plate Eq.(\ref{eq:weak-value}) is established, however in practice optical components are wavelength-dependent.
A quarter-wave plate with its optical axis at $45^{\cdot}$ direction and central angular frequency of $\omega_0$ can be described by a Jones matrix as:
\begin{equation}\label{eq:Uqwp}
\begin{array}{lll}
U_{\lambda/4} &= \frac{1}{2}\left(
  \begin{array}{cc}
    1 & -1 \\
    1 & 1\\
  \end{array}
\right)
\left(
  \begin{array}{cc}
    e^{-i\omega\tau_0} & 0 \\
    0 & e^{i\omega\tau_0} \\
  \end{array}
\right)
\left(
  \begin{array}{cc}
    1 & 1 \\
    -1 & 1\\
  \end{array}
\right)\\
&=
\left(
  \begin{array}{cc}
    \cos\omega\tau_0 & -i\sin\omega\tau_0 \\
    -i\sin\omega\tau_0 & \cos\omega\tau_0\\
  \end{array}
\right)
\end{array}
\end{equation}
where $\tau_0 = \frac{\pi}{4}/\omega_0$, and $\omega$ is the angular frequency of the light.

Denote the linear polarization state selected by PBS is $|\varphi_f\rangle$, the postselection of $|\varphi'_f\rangle$ can be divided as
\begin{equation}
\langle\varphi'_f|\hat{U}_{int}|\varphi_i\rangle|\phi\rangle
= (\langle\varphi_f|\hat{U}_{\lambda/4})\hat{U}_{int}|\varphi_i\rangle|\phi\rangle.
\end{equation}
Obviously, $|\varphi'_f\rangle = \hat{U}^{\dag}_{\lambda/4}|\varphi_f\rangle$ is frequency-dependent.
Combining with Eq.(\ref{eq:Uqwp}) and the following expressions, i.e.,
$$|\varphi_{f1}\rangle = \cos\gamma|H\rangle + \sin\gamma|V\rangle$$ and
$$|\varphi_{f2}\rangle = \sin\gamma|H\rangle - \cos\gamma|V\rangle,$$ we can derive
\begin{equation}
\begin{array}{lll}
|\varphi'_{f1}\rangle &=& (\cos\omega\tau_0\cos\gamma + i\sin\omega\tau_0\sin\gamma)|H\rangle \nonumber\\
&+& (\cos\omega\tau_0\sin\gamma + i\sin\omega\tau_0\cos\gamma)|V\rangle,
\end{array}
\end{equation}
and
\begin{equation}
\begin{array}{lll}
|\varphi'_{f2}\rangle &=& (\cos\omega\tau_0\sin\gamma - i\sin\omega\tau_0\cos\gamma)|H\rangle\nonumber\\
 &+& (-\cos\omega\tau_0\cos\gamma + i\sin\omega\tau_0\sin\gamma)|V\rangle.
\end{array}
\end{equation}
Consequently, the weak values can be calculated as
\begin{equation}
\begin{array}{lll}
A_{w1} &=& \frac{\langle\varphi'_{f1}|\hat{A}|\varphi_i\rangle}{\langle\varphi_{f1}|\varphi_i\rangle}\\
       &= & \frac{(\cos\omega\tau_0+i\sin\omega\tau_0)(\cos\gamma-\sin\gamma)}{(\cos\omega\tau_0-i\sin\omega\tau_0)(\cos\gamma+\sin\gamma)}\\
       &=& \frac{\cos\gamma-\sin\gamma}{\cos\gamma+\sin\gamma}e^{i2\omega\tau_0}\\
\end{array}
\end{equation}
\begin{equation}
\begin{array}{lll}
A_{w2} &=& \frac{\langle\varphi'_{f2}|\hat{A}|\varphi_i\rangle}{\langle\varphi_{f2}|\varphi_i\rangle}\\
       &=& \frac{(\cos\omega\tau_0+i\sin\omega\tau_0)(\cos\gamma+\sin\gamma)}{(\cos\omega\tau_0-i\sin\omega\tau_0)(\sin\gamma-\cos\gamma)}\\
       &=& \frac{\cos\gamma+\sin\gamma}{\sin\gamma-\cos\gamma}e^{i2\omega\tau_0}.
\end{array}
\end{equation}

In this case, $A_{wj}$ can not be taken outside of the integral of Eq.(\ref{eq:max-point}).
Compare the results shown in Fig.\ref{fig:jw-complete} and Fig.\ref{fig:wva-complete} we can see that in the balanced postselection scheme, Eq.(\ref{eq:first-order}) provides a good approximation, while in the unbalanced postselection scheme the wavelength-dependency of the quarter-wave plate involves an unnegligible deviation.

\end{document}